\documentclass[namedreferences]{solarphysics}

\usepackage[hyperref,optionalrh,showbiblabels]{spr-sola-addons} 
\usepackage{graphicx}        
\usepackage{color}           
\usepackage{breakurl}        

\chardef\us=`\_

\begin{document}

\begin{article}

\begin{opening}

\title{Determination of the Coronal and Interplanetary Magnetic-Field Strength and Radial Profiles from the Large-Scale Photospheric Magnetic Fields
}

%
\author{Irina~\surname{A.~Bilenko}
       }

%
 \runningauthor{I.A.~Bilenko}
 \runningtitle{A New Model for Coronal and Interplanetary Magnetic Field Determination}

%
  \institute{Moscow M.V. Lomonosov State University, Sternberg Astronomical
  Institute, Universitetsky pr.13, Moscow, 119234, Russia,
                     email: \url{bilenko@sai.msu.ru}    \\
           }

\begin{abstract}

A new model has been proposed for magnetic field determination at different
distances from the Sun during different solar cycle phases.
The model depends on the observed large-scale non-polar ($\pm55^{\circ}$)
photospheric magnetic fields and that measured at polar regions
from $55^{\circ}$ N to $90^{\circ}$ N and
from $55^{\circ}$ S to $90^{\circ}$ S, which are the visible manifestations of cyclic changes in the
toroidal and poloidal components of the global magnetic field of the Sun.
The modeled magnetic field is determined as the superposition of the non-polar
and the polar photospheric magnetic field  cycle variations.
The agreement between the model predictions and magnetic fields derived from direct,
in-situ, measurements at different distances from the
Sun, obtained by different methods, and at different solar activity phases is quite satisfactory.
From a comparison of the magnetic fields as observed and as calculated from the
model at 1 AU, it should be concluded that the model magnetic-field variations
adequately explains the major features of the IMF $B_x$ component cycle evolution at the Earth's orbit.
The model CR-averaged magnetic fields correlate with CR-averaged IMF $B_x$ component at the  Earth's orbit
with a coefficient of 0.688, and for seven CR-averaged data the correlation reaches 0.808.
The model magnetic-field radial profiles were compared with that of the already existing models.
In contrast to existing models our model provides realistic magnetic-field radial
distributions over a wide range of heliospheric distances at different cycle phases taking
into account the cycle variations of the solar toroidal and poloidal magnetic fields.
The model should be regarded as a good approximation
of the cycle behavior of the magnetic field in the heliosphere.
In addition, the decrease in the non-polar and polar photospheric magnetic fields
has been revealed. Both magnetic fields during solar cycle maxima and that
during minima phases decreased from Cycle 21 to Cycle 24. It means that
both the toroidal and poloidal components and therefore, the solar global magnetic field
decreased from Cycle 21 to Cycle 24.

\end{abstract}

%
\keywords{Magnetic fields, Corona; Magnetic fields, Interplanetary; Magnetic fields, Models; Solar Cycle}

\end{opening}

%
 \section{Introduction}  \label{intro}

Solar magnetic fields are swept into interplanetary space by the solar wind flows.
Solar fields dominate the structure and dynamics of the heliosphere.
The heliosphere is spatially and temporally varying with respect to the magnetic field.
Therefore, the regularity in magnetic-field space and time distribution at different distances
from the Sun and during different solar cycle phases are not yet well known.
This is largely due to the weakness of coronal magnetic fields
especially at distances greater than 5 Rs (solar radii).
So, to analyze different processes in the solar corona and interplanetary space,
the magnetic-field distribution and cycle evolution must be ascertained.

Currently a number of different methods are used to measure the
coronal magnetic field at different distances from the Sun
(\opencite{Akhmedov1982}; \opencite{Lin2000}, \opencite{Lin2004},
\opencite{Bogod2016}; \opencite{Gelfreikh1987}).
Some measurements of coronal magnetic fields were made at different
wavelength (\opencite{Lin2000}; \opencite{Lin2004}; \opencite{Raouafi2016}).
Faraday rotation measurement technique is commonly used in estimating the coronal
magnetic-field strengths within 10 Rs (\opencite{Patzold1987}; \opencite{Sakurai1994};
\opencite{Spangler2005}; \opencite{Ingleby2007}). \cite{Patzold1987} found that
the coronal magnetic field at 5 Rs is around
100$\pm$50 mG, \cite{Sakurai1994} derived magnetic field at 9 Rs as 12.5$\pm$2.3 mG,
and \cite{Spangler2005} found a value of 39 mG at 6.2 Rs.
\cite{Ingleby2007} measured Faraday rotation with the VLA at frequencies of 1465 and 1665 MHz
and found that the coronal magnetic field is in the range of 46\,--\,120 mG at heliocentric distance of 5 Rs.
\cite{Xiong2013} used coordinated observations in polarized white light and Faraday
rotation measurements to determine the spatial position and magnetic field of an interplanetary sheath.
Magnetic field can also be derived from the measurements of the solar wind plasma
using pulsars (\opencite{Ord2007}; \opencite{You2012}).

Several methods of magnetic field determination were developed using solar radio emission.
Using data from SOHO/UVCS and radio spectrograph observations, \cite{Mancuso2003}
estimated magnetic field and plasma properties in active region corona.
The results show that the magnetic field
is expressed by the inequality $B(r) \le (0.6 \pm 0.3)(r -1)^{-1.2}$ G,
that is valid in the range $1.5 \le r \le 2.3$ Rs  (\opencite{Mancuso2003}).
Magnetic field strength can be derived from the band splittings
in type II radio bursts, if the coronal density distribution is given
(\opencite{Vrsnak2002}; \opencite{Cho2007}; \opencite{Hariharan2014}).
Using band splitting of coronal type II radio bursts,
\cite{Cho2007} obtained a coronal magnetic-field
strength of 1.3\,--\,0.4 G in the height range of 1.5\,--\,2 Rs.

Recent investigations demonstrate that shock waves
propagating into the corona and interplanetary space,
associated with major solar eruptions, can
be used to derive the strength of magnetic fields
over a very large interval of heliocentric distances and latitudes.
So-called 'standoff-distance' method was developed and is often used now
(\opencite{Gopalswamy2011}; \opencite{Gopalswamy2012}; \opencite{Kim2012};
\opencite{Bemporad2016}; \opencite{Schmidt2016}).
In this method, a coronal mass ejection (CME) and CME-driving shock dynamics are analyzed.
The shock standoff distance, speed, and the radius of CME flux rope curvature
are measured. Alfv\'en speed and Mach number can be derived using the measured data.
Then the magnetic field can be derived applying some of a coronal density model.
Using this method, \cite{Gopalswamy2011} have found that
the magnetic field declines from 48 to 8 mG in the distance range from 6 to 23 Rs.
Following the standoff-distance method and using data from
Coronagraph 2 and Heliospheric Imager I instruments on board
the Solar Terrestrial Relations Observatory, \cite{Gopalswamy2012}
have found that the radial magnetic field strength decreases from
28 mG at 6 Rs to 0.17 mG at 120 Rs. They also noted that the radial profile
of magnetic-field strength can be described by a power
law.
\cite{Kim2012} performed a statistical study by applying this method to
10 fast ($\ge$1000 km s$^{-1}$) limb CMEs (LASCO data),
to measure the magnetic-field strength in the solar corona in the height
range 3\,--\,15 Rs. They found that the magnetic-field strength is in the range 6\,--\,105 mG.
They show, that the magnetic-field values derived with the standoff-distance method
are consistent with other estimates in a similar distance range.

But despite the increasing number of space missions and
despite the progress made recently in the solar and interplanetary
space observations, the reliable measurements of the coronal
and interplanetary magnetic-field strength and orientation at different
distances and cycle phases do not exist.
At present, magnetic fields are routinely measured in the photospheric level and
NSO SOLIS/VSM also observes the full-disk chromospheric field using Ca II 8542 nm line,
but not in the solar corona and IMF.
So, the magnetic field in the solar corona and interplanetary space is estimated
from the observed photospheric fields using different extrapolation
techniques into the solar corona. Coronal magnetic fields are investigated
and modeled at differen distances from the Sun up to several Rs.
Some models at coronal heights are based on magnetic fields in active regions
(\opencite{Brosius2006}; \opencite{Bogod2008}, \opencite{Bogod2012}, \opencite{Kaltman2012}).
Different models are developed both analytic (\opencite{Banaszkiewicz1998})
and numerical including MHD-models, magnetohydrostatics, force-free or
potential-field models (\opencite{Gibson1995}; \opencite{Wiegelmann2004};
\opencite{Wiegelmann2017}, and references therein).

It should also be noted, that the majority of models describe the
magnetic-field distribution in radial direction from the Sun only.
As a rule, in such models the measurements of magnetic field obtained at different times are
summarized in one curve regardless of a cycle or a cycle phase. However,
it is well known that the magnetic field in the quiet corona during sunspot minimum
is much lower than  that determined for an average sunspot maximum and that
spherical symmetry is not observed. Furthermore, at solar maximum, the magnetic
field is different in different solar cycles, as a result of different level of activity.
Below the height of  $\approx$ 3 Rs the magnetic
field is governed by the active-region fields. Above height of $\approx$ 3 Rs the
radial field, decreasing as $R^{-2}$, becomes dominant.
It should be noted, that all coronal magnetic-field models, describing the magnetic fields
above $\approx$ 3 Rs give only one value for the distance required
regardless of the cycle phase.
It is necessary to use more realistic magnetic-field radial distribution models taking
into account the solar magnetic-field cycle variations.
Our attention thus has been directed to a detailed description of
the observed solar magnetic-field distribution and cycle variation, to
create a model of magnetic-field radial distribution and cycle variations
from 1 Rs to 1AU. We then compare the magnetic fields
derived using our model with that measured at different time and distances,
as well as with magnetic-field profiles from already existing models.
The article will summarize most recent models and results on magnetic field measurements.

The paper is organized as follows.
The data are described in Section \ref{secdata}.
In Section \ref{secmag}, the photospheric and interplanetary magnetic-field
distribution and cycle evolution are presented, and a new model of magnetic field calculation
at different distances from the Sun with the consideration of the solar cycle magnetic-field variations
is suggested. The magnetic fields measured by different methods
are compared with that calculated using our model in Section~\ref{secmagobs}.
The comparison of our model calculated results with that derived using already existing models is made in Section~\ref{secmagmod}.
The results are discussed in Section~\ref{secdiscussion}.
The main results are listed in Section~\ref{secconclusion}.

\section{Data}   \label{secdata}

Data on the large-scale photospheric magnetic fields
from the Wilcox Solar Observatory (WSO) were used for the years 1976\,--\,2015.
Full-disk synoptic maps span a full Carrington Rotation (1 CR = 27.2753 days).
They are assembled from individual magnetograms observed during a solar rotation.
WSO synoptic maps only represent the radial component of the photospheric field
(derived from observations of the line-of-sight field component by assuming the field
to be approximately radial).
The entire data set consists of 530 synoptic maps and covers CRs
1642\,--\,2172 (June 1976\,--\,December 2015).
Synoptic map magnetic-field data consist of $30$ data points in equal steps
of sine latitude from $+70^{\circ}$ to $-70^{\circ}$. As the solar magnetic fields are
measured from the Earth, the field above $70^{\circ}$ in the North and South hemispheres is not resolved. Longitude is presented in $5^{\circ}$ intervals (\opencite{Duvall1977}; \opencite{Hoeksema1986}).

WSO polar field observations were used in this study.
The Sun's polar magnetic-field strength is measured in the polemost 3' apertures at WSO each day in the North and South hemispheres. The line-of-sight magnetic field between about $\pm55^{\circ}$ and the pole in the corresponding hemisphere is  measured. The daily polar field measurements are averaged each 10 days in a centered 30-day window.
The solar coordinates of the apertures shift and the square aperture at the pole is oriented differently on the Sun during each measurement due to the Earth movement above and below the equator each year.

Data on interplanetary magnetic field (IMF) were obtained from multi-source OMNI 2 data base.
(\opencite{King2005}). From the OMNI 2 data base the hourly mean values of the IMF measured by various
spacecraft near the  Earth's orbit were considered.
Only the $B_x$ component of the OMNI 2 IMF was used in this study.

\section{Large-Scale Photospheric and Coronal Magnetic Field Variations in Cycles 21\--\,24}  \label{secmag}

In order to study the variations of the magnetic-field distribution
in the solar corona, we have used the direct observations of
the large-scale non-polar ($\pm55^{\circ}$) and
polar (from $55^{\circ}$ N to $90^{\circ}$ N and
from $55^{\circ}$ S to $90^{\circ}$ S) photospheric magnetic fields
which are the visible manifestations of the cyclic changes in the
toroidal and poloidal components of the global magnetic field of the Sun.
In  addition to the general magnetic field distribution, occasional
displacements and oscillations appear in localized regions of the solar corona
due to magnetic fields carried out by CMEs, flows from coronal holes, flares, ets.
Therefore, the CR averaged data were used.

\begin{figure}
   \centerline{\includegraphics[width=1.\textwidth,clip=]{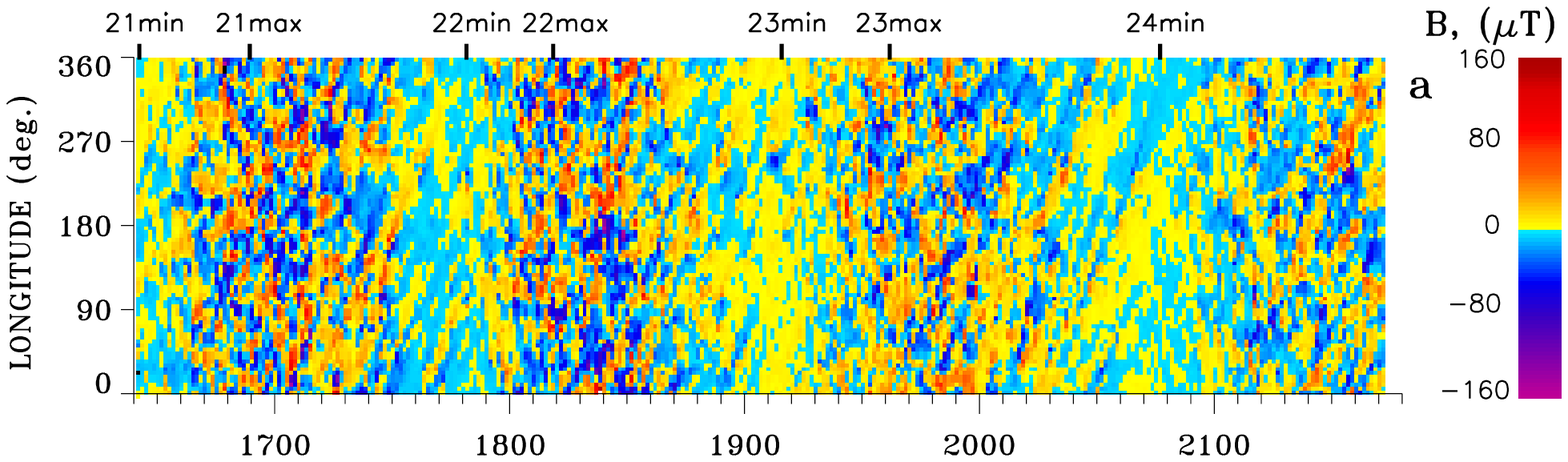}}
\vspace{-0.8cm}
   \centerline{\includegraphics[width=1.\textwidth,clip=]{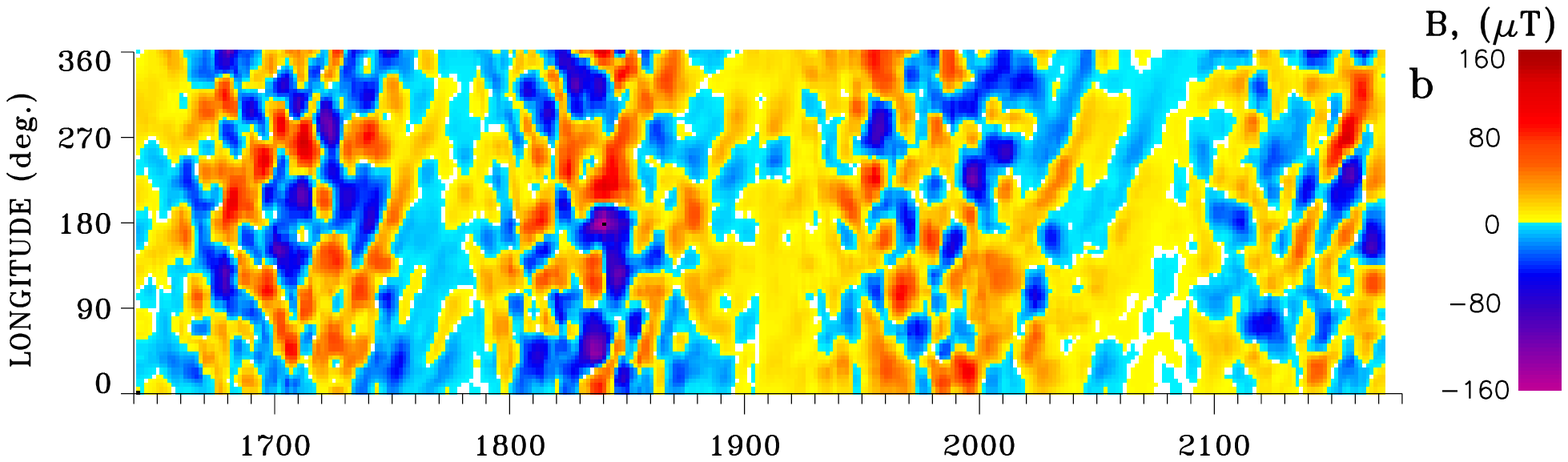}}
\vspace{-0.8cm}
   \centerline{\includegraphics[width=1.\textwidth,clip=]{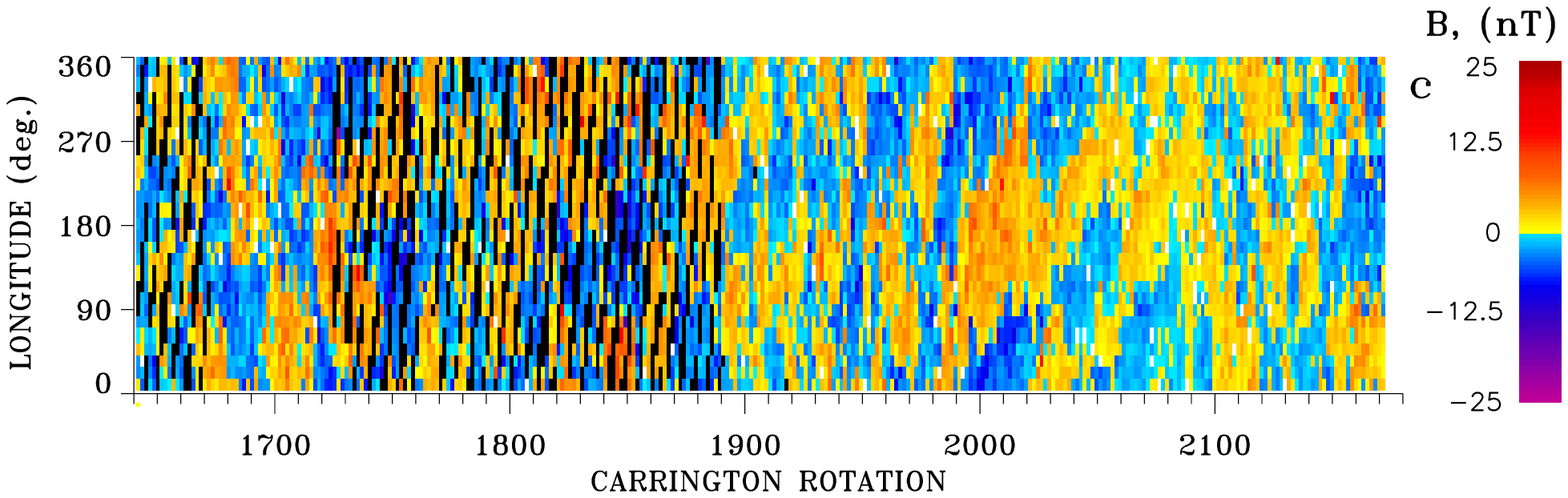}}
\caption{Longitudinal diagrams.
            (a) Large-scale photospheric magnetic fields.
            (b) Large-scale photospheric magnetic fields smoothed by 7x7 CRs.
            (c)  IMF at 1 AU.
               Red denotes the positive-polarity magnetic fields,
               blue denotes the negative-polarity magnetic fields.
               Black denotes the missing data.
               The maxima and minima of Cycles 21\,--\,24 are marked at the top.}
   \label{maglon}
\end{figure}

For the description of the non-polar magnetic field cycle variations,
we created diagrams of the photospheric magnetic-field distribution based on
the observed large-scale photospheric magnetic fields
from $55^{\circ}$S to $55^{\circ}$N latitude through Cycles 21\,--\,24.
Figure~\ref{maglon}a shows the longitudinal time-space distribution
of the large-scale photospheric magnetic fields.
The smoothed photospheric magnetic-field longitudinal
distribution is presented in Figure~\ref{maglon}b.
The CR-averaged distribution of positive- and negative-polarity
interplanetary magnetic fields (IMF) at the Earth's orbit is shown in Figure~\ref{maglon}c.
The longitudinal diagrams were created in a CR-rotation system.
The $x$-axes denote the date of 0$^{\circ}$ CR longitude at the central
meridian, and the $y$-axes denote longitude in each CR magnetic-field diagram.
The detailed description of the magnetic-field longitudinal diagram creation and
solar global magnetic field evolution is given in \cite{Bilenko2012}, \cite{Bilenko2014}.
The maxima and minima of the cycles are marked at the top of Figure~\ref{maglon}.
Figure~\ref{maglon} demonstrates the close connection of the magnetic field direction
and strength variations of the IMF at the Earth's
orbit and the photospheric magnetic fields.

\begin{figure}
   \centerline{\includegraphics[width=1.\textwidth,clip=]{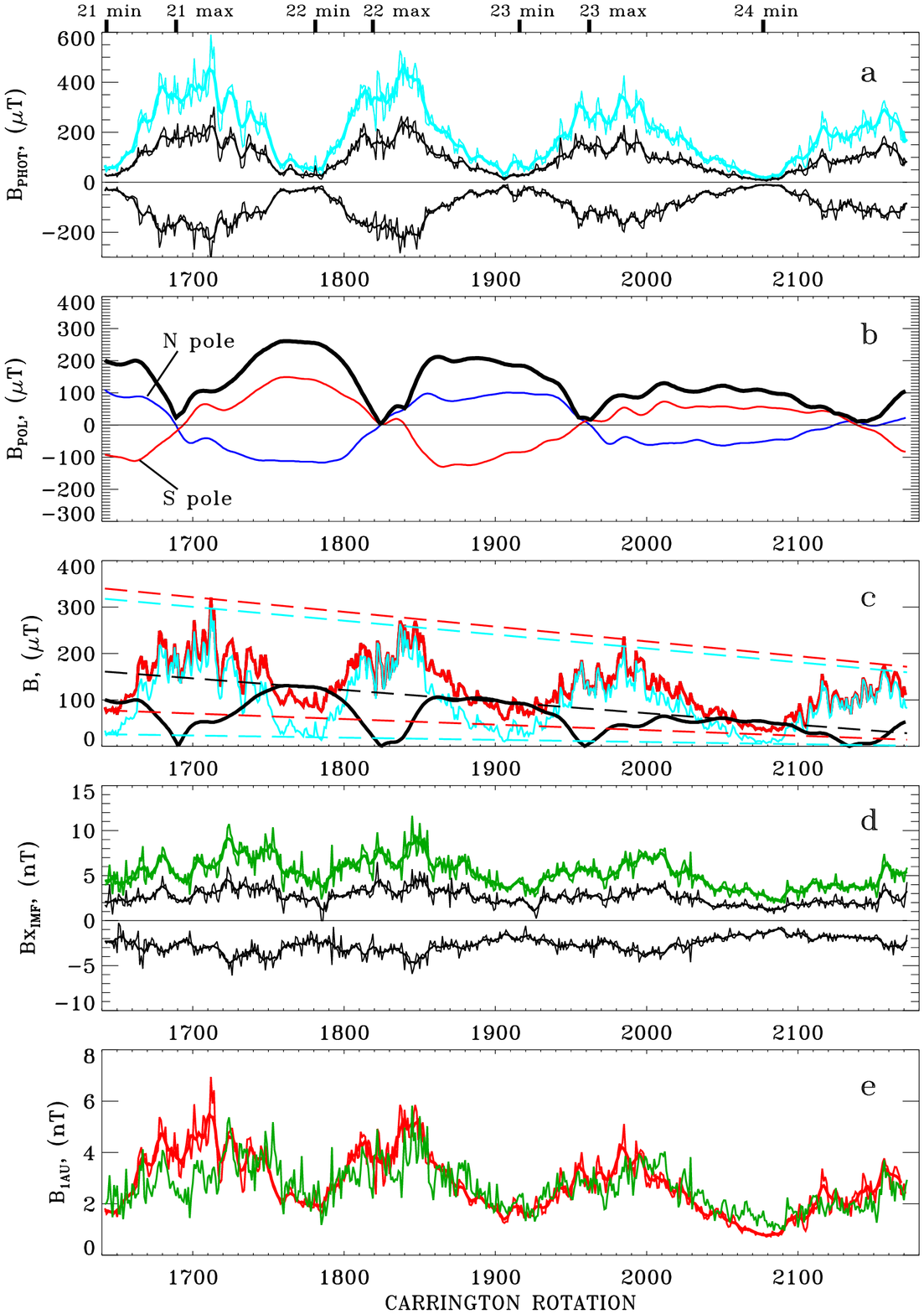}}
      \caption{(a) Photospheric non-polar magnetic fields from the longitudinal diagram Figure~\ref{maglon}a.
         Black denotes the positive- and negative-polarity photospheric magnetic fields,
         and light blue denotes the sum of their moduli.
      (b) CR-averaged polar magnetic-field cycle variations at the North (N) and South (S) poles,
          and the sum of their moduli (black line).
      (c) Sum of the positive- and negative-polarity photospheric (non-polar) magnetic
         fields (light blue line) from (a) and the sum of the North and South
         pole magnetic fields (black line) from (b).
         Red line denotes the sum of polar and non-polar magnetic fields.
      (d) IMF measured at 1 AU.
         Black denotes the CR-averaged positive- and negative-polarity magnetic fields,
         and green denotes the sum of their moduli.
      (e) Red line denotes magnetic field at 1 AU calculated from Eq.~\ref{fmag} as the sum the
         non-polar and polar magnetic fields.
         Green line denotes the half of the sum of CR-averaged positive- and negative-polarity IMF
          at 1 AU.
          In (a), (c), (d), and (e) thin lines correspond to CR-averaged and thick lines to seven CR-averaged data.
         The maxima and minima of Cycles 21\,--\,24 are marked at the top.}
   \label{mag}
 \end{figure}

Figure~\ref{mag}a shows the cycle changes of the
photospheric large-scale magnetic field obtained from the longitudinal distribution of the photospheric magnetic
fields by averaging over the latitude from 55 N to 55 S for every CR displayed in Figure~\ref{maglon}a.
Black denotes the positive- and negative-polarity photospheric magnetic fields,
and light blue denotes the sum of their moduli.
Thin lines show changes in the CR-averaged magnetic fields and
thick lines correspond to the seven CR-averaged data.
The maxima and minima of Cycles 21\,--\,24 are marked at the top of Figure~\ref{mag}.
Examination of Figures~\ref{maglon} and \ref{mag} shows that the magnetic
field does not change smoothly from the minimum of
solar activity to maximum, but in the form of some impulses.
These changes reflect cyclical changes in the structure and strength
of the solar global magnetic field (\opencite{Bilenko2012}; \opencite{Bilenko2014}; \opencite{Bilenko2016}).
The maximum of magnetic field magnitudes decreased and it was
the highest in Cycle 21 and the lowest in Cycle 24. The magnetic field strength
decrease can be given by

\begin{equation}
B_{max}(t) = 317.82 - 0.30 \times t_{CR}
\label{fphotmax}
\end{equation}

It should be noted, that the minimum values of the magnetic field during
solar activity minima also decreased from Cycle 21 to Cycle 24,
and it was also the highest in Cycle 21 and the lowest in Cycle 24.
We find a regression line

\begin{equation}
B_{min}(t)= 26.15 - 0.47 \times t_{CR}
\label{fphotmin}
\end{equation}

Such cyclic changes in magnetic fields (Figure~\ref{mag}a), reflect the
time evolution of the toroidal component of the solar global magnetic field. It means that
the toroidal component decreases from Cycle 21 to Cycle 24.

Magnetic-field cycle variations at the solar poles
are the observational manifestation of the poloidal component of the
solar global magnetic field.
Figure~\ref{mag}b shows CR-averaged magnetic-field variations
in the North (blue line) and South (red line) poles. Black line denotes the sum of their moduli.
Polar magnetic fields at the North and the South poles diminished
from Cycle 21 to Cycle 24. The cyclic changes in magnetic field (Figure~\ref{mag}b) reflect the
time evolution of the poloidal component of the solar global magnetic field
from Cycle 21 to Cycle 24.
It means that both components have decreased from Cycle 21 to Cycle 24.
The decrease of the polar magnetic field can be expressed by

\begin{equation}
B_{pol}(t) = 116.04 - 0.25 \times t_{CR}
\label{fpolmin}
\end{equation}

The magnetic field in the ecliptic plane can be modeled as the superposition
of toroidal and poloidal component cycle variations.
The toroidal component dominates at solar maxima and the influence of the
poloidal component increases during solar minima.
The toroidal component increases during solar maxima because of the enhancement of the
active-region magnetic fields.
At solar minima, about 60\% of the solar sphere above 2 Rs is dominated by polar coronal holes.
So, the total magnetic-field strength can be determined as the sum of the non-polar photospheric
magnetic fields (toroidal component of the solar global magnetic field, Figure~\ref{mag}a)
and the polar magnetic field (poloidal component) obtained from Figure~\ref{mag}b.
Both non-polar (light blue line) and polar (black line) magnetic-field components
and their sum (red line) are shown in Figure~\ref{mag}c.
We detect the magnetic-field components that directly influence on
the measured IMF at the Earth's orbit.
It is assumed that the corona is in a steady state from 1 Rs to 1 AU.
It should be noted, that both non-polar and polar fields are observed in
ecliptic plane from the Earth.
So, we can derive the radial component ($B_x$) of the IMF from the observed
non-polar and polar photospheric magnetic-field strength in ecliptic plane using a relation

\begin{equation}
B(t,r) = \left(\frac{B_{phot+}(t) + B_{phot-}(t)}{2} + \frac{B_{pol_N}(t)+B_{pol_S}(t)}{2}\right) \times \left(\frac{1}{r}\right)^2
\label{fmag}
\end{equation}

\noindent where $B_{phot+}$ and $B_{phot-}$ are the absolute values of the positive- and negative-polarity magnetic fields at the photosphere derived from
Figure~\ref{maglon}a and shown in Figure~\ref{mag}a for each CR,
and they represent the toroidal component of the
solar magnetic field; $B_{pol_N}$ is the absolute value of the observed North polar magnetic field
and $B_{pol_S}$ is the absolute value of the South polar magnetic field (Figure~\ref{mag}b).
They represent the poloidal component of the
solar global magnetic field; $r$ is the distance from the center of the Sun in units of the solar radius.
The different behavior of polar and non-polar magnetic fields is also known from active region (\opencite{Zharkov2007}) and  coronal hole (\opencite{Lowder2017}; \opencite{Bilenko2016}) cycle evolution. Polar and non-polar magnetic fields represent the different components of the solar global magnetic field. Their cycle evolution is different. Therefore, they were included in Equation~\ref{fmag} as separate components.

As noted above, the magnitudes of the toroidal and poloidal magnetic-field
components of the solar global magnetic field decreased from Cycle 21 to Cycle 24.
We find a regression line for our model (Eq.~\ref{fmag}) magnetic-field cycle maxima

\begin{equation}
B_{m_{max}}(t) = 339.93 - 0.32 \times t_{CR}
\label{fmagmax}
\end{equation}

an expression for the minima can be written as

\begin{equation}
B_{m_{min}}(t) = 77.67 - 0.12 \times t_{CR}
\label{fmagmin}
\end{equation}

In Figure~\ref{mag}c, the dashed lines are the regression lines derived in
Equations~\ref{fphotmax}, \ref{fphotmin}, \ref{fpolmin},
\ref{fmagmax}, \ref{fmagmin}.

In Figure~\ref{mag}d, the variation of the observed $B_x$ component of the
IMF at the Earth's orbit is shown.
Black denotes the CR-averaged positive- and negative-polarity magnetic fields,
and green denotes the sum of their moduli. Thin lines show CR-averaged IMF
and  thick lines correspond to the seven CR-averaged data.

\begin{figure}
   \centerline{\includegraphics[width=1.\textwidth,clip=]{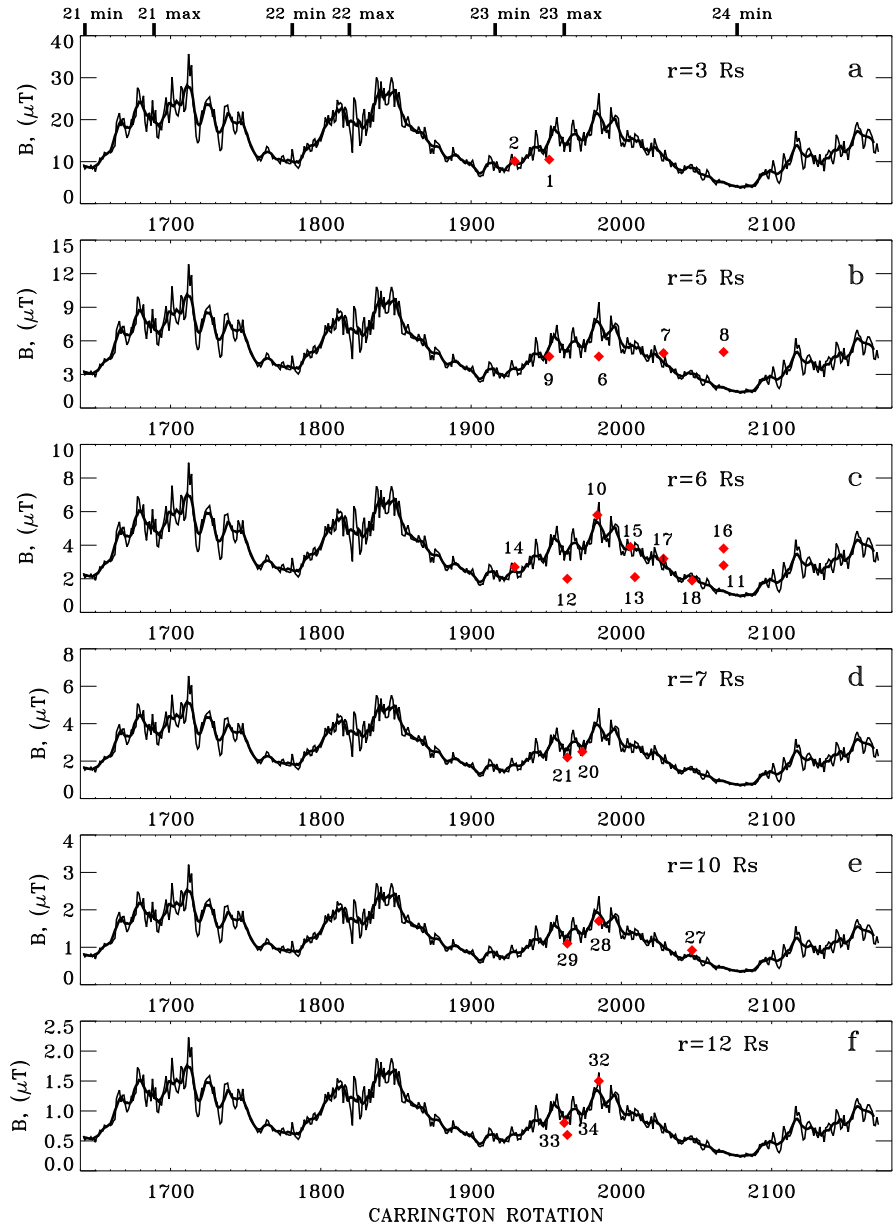}}
      \caption{Model calculated magnetic fields at 3 Rs, 5 Rs, 6 Rs, 7 Rs, 10 Rs, 12 Rs.
      Red points denote the directly measured IMF (from Table~\ref{tab1}) that are close to these distances.
      The labels of the points correspond to their number in Table~\ref{tab1}.
     The maxima and minima of Cycles 21\,--\,24 are marked at the top.}
   \label{magexmp}
 \end{figure}

 \begin{figure}
   \centerline{\includegraphics[width=1.\textwidth,clip=]{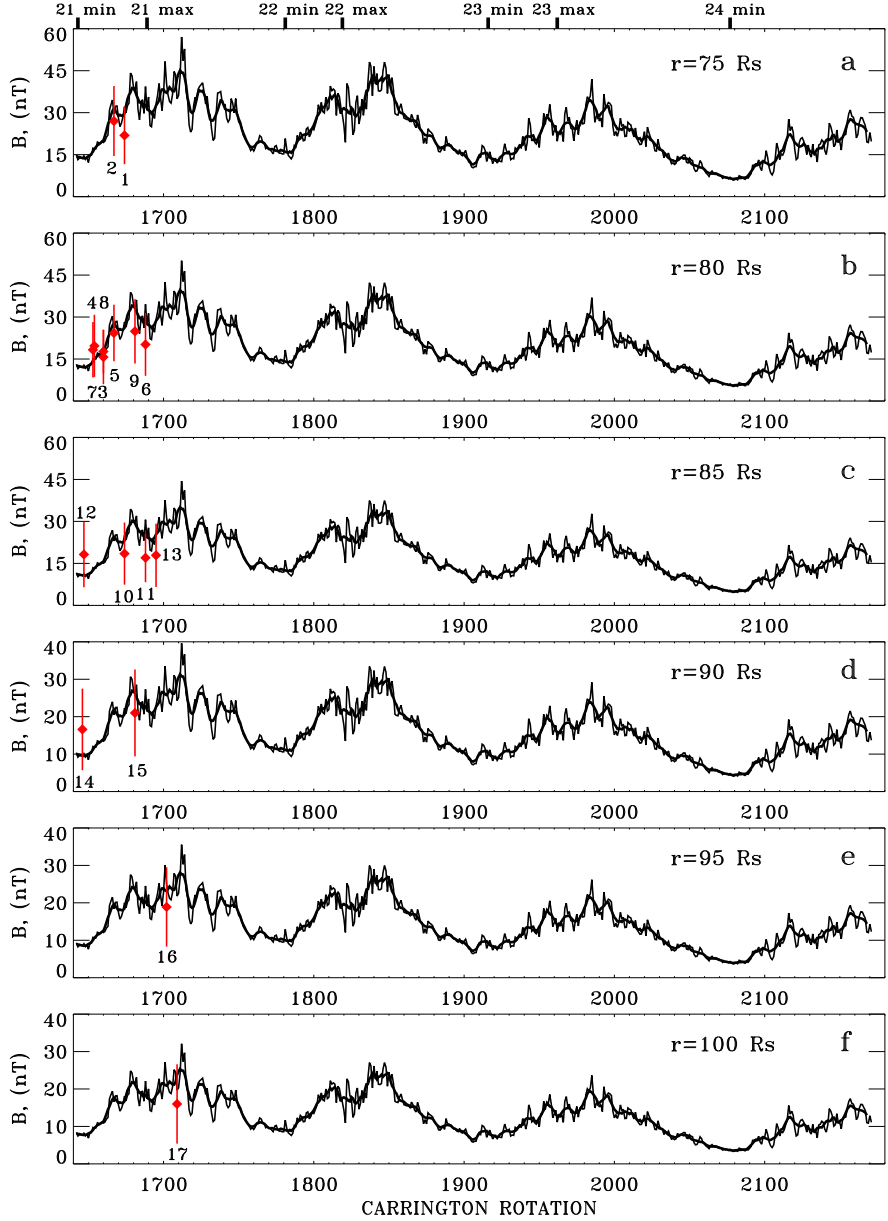}}
      \caption{Model calculated magnetic fields at 75 Rs, 80 Rs, 85 Rs, 90 Rs, 95 Rs, 100 Rs.
      Red points denote the IMF measured by Helios 1 and Helios 2 spacecraft from Table~\ref{tab2}.
      The labels of the points correspond to their number in Table~\ref{tab2}.
      The maxima and minima of Cycles 21\,--\,24 are marked at the top.}
   \label{maghelios}
 \end{figure}

 \begin{figure}
   \centerline{\includegraphics[width=1.\textwidth,clip=]{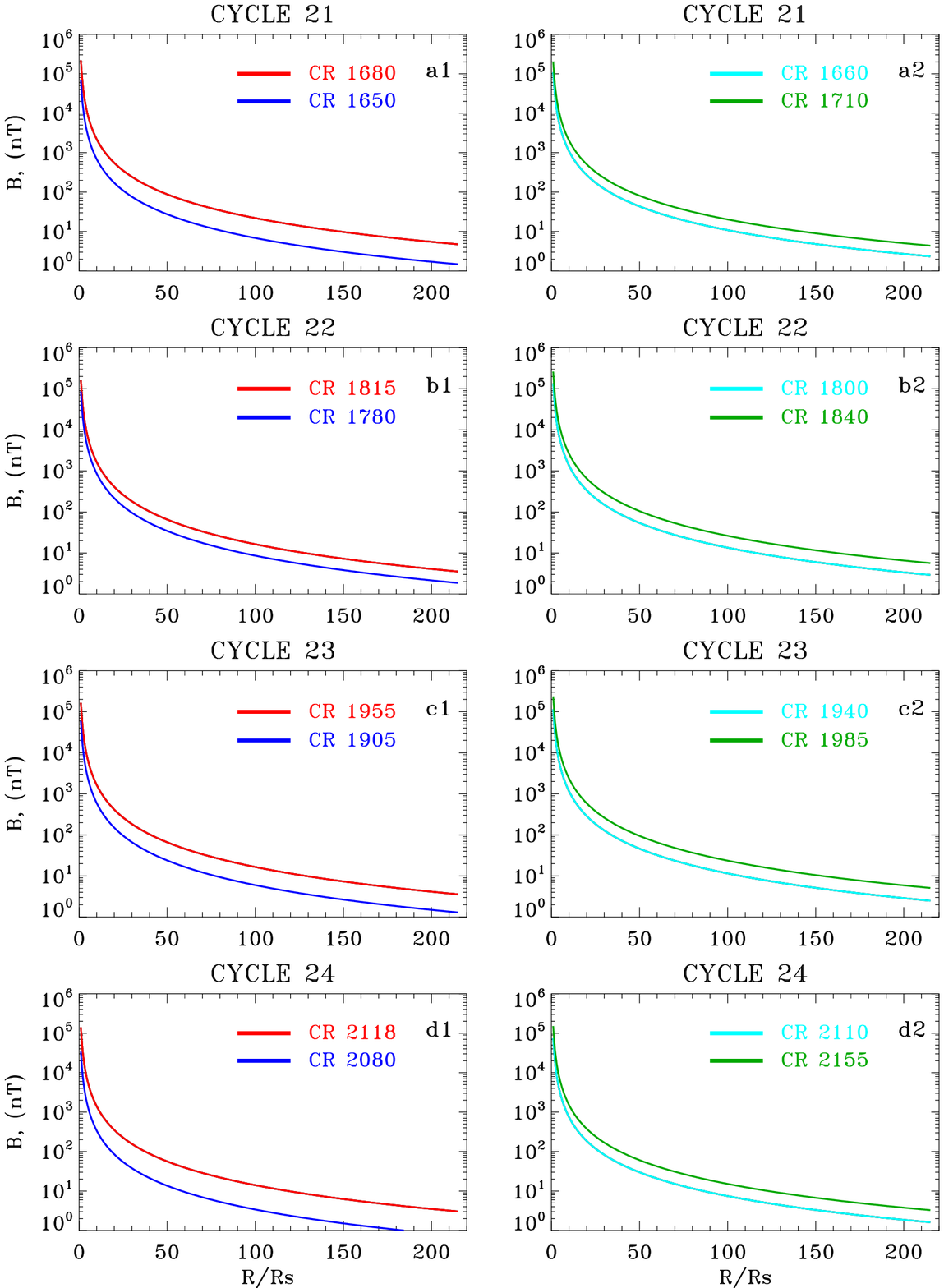}}
      \caption{Magnetic field distributions calculated using our model (Eq.~\ref{fmag})
       for different corona activity states.
       In each panel, the certain CRs are shown for different phases of solar activity.
Blue profiles denote solar activity minima corona; red profiles correspond to the solar activity maxima corona;
light blue profiles denote solar corona at the rising phases; green profiles denote solar corona at the declining phases.}
   \label{magr}
 \end{figure}

In Figure~\ref{mag}e, the magnetic field derived
at 1 AU using our model (Eq.~\ref{fmag}) is shown in red color.
Green line denotes the half of a sum of CR-averaged positive- and negative-polarity
IMF magnetic fields at 1 AU in Cycles 21\,--\,24.
CR-averaged magnetic fields were compared. Both magnetic field retrieved from the photospheric magnetic fields and IMF measured at the Earth's orbit was calculated the same way. In Eq.~\ref{fmag}, the  magnetic fields were used as the half of the sum of the moduli of positive- and negative-polarity non-polar and polar magnetic fields averaged during each CR. Likewise IMF is also the sum of CR-averaged of the moduli of positive- and negative-polarity $B_x$ component of the IMF divided by 2.
As seen from Figure~\ref{mag}e, our model fits $B_x$ component of IMF at 1 AU very
well particularly during Cycles 23 and 24.
From a comparison of the magnetic fields as observed and as calculated from our
model, we conclude that our model magnetic-field strength variation adequately explains
the major features of the IMF changes during Cycles 21\,--\,24.
However, there are certain differences.
The difference is greater during the maximum phase in Cycle 21,
but  the mismatch is significantly less in the maxima of Cycles 22 and 24 (Figure~\ref{mag}e).
Some peaks in magnetic fields derived using our model (Equation~\ref{fmag})
coincide with peaks in IMF and some peaks follow the peaks in the IMF.
The coincidence between IMF and our model predicted magnetic fields at 1 AU
is more pronounced during the declining phases.
The reduction of the misalignment between the interplanetary magnetic field and that
predicted using our model observed from Cycle 21 to Cycle 24, can be explained by the
fact that the quality of observations and measurements of both
IMF and solar magnetic fields has significantly improved by now.
We have also carried out correlation analysis of our results in order to
quantify agreement. The correlation coefficients between the observed
IMF and magnetic fields at 1 AU predicted by our model were calculated.
The model CR-averaged calculated magnetic fields correlate with IMF at the  Earth's orbit
with a coefficient of 0.688.
Correlation between seven CR-averaged calculated magnetic fields with seven CR-averaged $B_x$ component of the IMF reaches 0.808.
For Cycles 21\,--\,24, the model CR-averaged calculated magnetic fields
correlate with CR-averaged $B_x$ component of the IMF at the  Earth's orbit
with a coefficients of 0.42, 0,69, 0.73, and 0.74 and that for seven CR-averaged values
with a coefficients of 0.52, 0,88, 0.85, and 0.88
Again, this implies a very good correlation.
So, our model, the Equation~\ref{fmag}, is a good representation of the
measured radial ($B_x$) component of the IMF in the ecliptic plane.

Figures~\ref{magexmp}, \ref{maghelios} show some examples of the model
magnetic-field evolution during Cycles 21\,--\,24
calculated at distances of 3 Rs, 5 Rs, 6 Rs, 7 Rs, 10 Rs, and 12 Rs (Figure~\ref{magexmp})
and 75 Rs, 80 Rs, 85 Rs, 90 Rs, 95 Rs, 100 Rs (Figure~\ref{maghelios}).
From Figures~\ref{magexmp}, \ref{maghelios}, it is seen that the magnetic field declines
from $\approx 20 \mu T $ to $\approx 20 nT$ in the distance range from 3 Rs to 100 Rs.
The magnetic-field cycle variations derived for different distances
and illustrated in Figures~\ref{magexmp} and \ref{maghelios} should be regarded as a
good approximation of the cycle behavior of the radial magnetic-field profiles in the
heliosphere. Thus, this method of the magnetic-field calculation is a useful tool for
the description of solar corona and interplanetary large-scale magnetic-field cycle evolution
at differen distances from the Sun.

Using our model, we can also derive the radial magnetic-field strength profiles  for
different solar-activity phases and for different distances from the Sun.
Figure~\ref{magr} shows magnetic-field radial profiles from 1 Rs to 1 AU. The  profiles
were derived using Equation~\ref{fmag} as a function of distance for
different phases of Cycles 21\,--\,24.
The magnetic-field profiles were calculated for corona activity states:
(1) blue lines denote solar-activity minima corona; (2) red lines correspond to the solar activity maxima corona;
(3) light blue lines denote the solar corona at the rising phases; (3) green lines denote the solar corona at the declining phases. In each panel, the certain CRs are shown for different phases of solar activity.

\section{Coronal Magnetic Field Measurements in Cycles 21\,--\,24}  \label{secmagobs}

Table~\ref{tab1} summarizes the observed magnetic fields (column 6) determined by
different authors at different distances (column 4) from the Sun
and those calculated using our model, Equation~\ref{fmag}, (column 5)
at the same distances and the same time (columns 2, 3).
Magnetic-field measured data were taken from the articles cited in column 7.

\begin{table}
\caption{Magnetic fields observed and that calculated using
                our model at the same distances from the Sun and the same time.}
\label{tab1}
\tabcolsep=0.6em
\begin{tabular}{llccccl}
\hline
  N & Date & CR &  R   &      B       &      B      &   References    \\
     &         &       & (Rs)&    calc.    &    obs.    &         \\
     &        &        &       & ($\mu$T) & ($\mu$T)&          \\
\hline
   1 &  2   &   3   &    4  &      5       &       6      &       7   \\
\hline
  1  &  25--Jul--1999  &  1952  &  3.08  &  13.99  & 10.5    &   Kim et al. (2012) \\
  2  &  14--Nov--1997  &  1929  &  3.33  &   7.86  & 10.1    &   Kim et al. (2012) \\
  3  &  24--Oct--2003  &  2009  &  3.56  &  11.79  &  5.6    &   Kim et al. (2012) \\
  4  &   1--Apr--2001  &  1974  &  4.21  &   7.70  &  5.2    &   Kim et al. (2012) \\
  5  &  22--Mar--2002  &  1987  &  4.3   &   8.09  &  1.9    &   Bemporad et al. (2010) \\
  6  &  14--Jan--2002  &  1985  &  4.82  &  10.17  &  4.6    &   Kim et al. (2012) \\
  7  &  Mar-Apr--2005  &  2028  &  5.0   &   3.66  & 4.6-5.2 &   Ingleby et al. (2007) \\
  8  &  25--Mar--2008  &  2068  &  5.0   &   1.87  &  5.0    &   Gopalswamy et al. (2011) \\
  9  &  25--Jul--1999  &  1952  &  5.55  &   4.31  &  4.6    &   Kim et al. (2012) \\
  10 &  28--Dec--2001  &  1984  &  5.98  &   5.66  &  5.8    &   Kim et al. (2012) \\
  11 &  5--Apr--2008   &  2068  &  6.0   &   1.3   &  2.8    &   Poomvises et al. (2012) \\
  12 &  15--Jun--2000  &  1964  &  6.06  &   3.42  &  2.0    &   Kim et al. (2012) \\
  13 &  24--Oct--2003  &  2009  &  6.10  &   4.02  &  2.1    &   Kim et al. (2012) \\
  14 &  14--Nov--1997  &  1929  &  6.19  &   2.27  &  2.7    &   Kim et al. (2012) \\
  15 &  16--Aug--2003  &  2006  &  6.2   &   3.75  &  3.9    &   Spangler et al. (2005) \\
  16 &  25--Mar--2008  &  2068  &  6.2   &   1.22  &  3.8    &   Gopalswamy et al. (2011) \\
  17 &  Mar--Apr--2005 &  2028  &  6.2   &   2.38  & 3.0-3.4 &   Ingleby et al. (2007) \\
  18 &  28--Aug--2006  &  2047  &  6.2   &   2.17  &  1.9    &   You et al. (2012) \\
  19 &  5--May--2000   &  1962  &  6.35  &   2.76  &  3.5    &   Kim et al. (2012) \\
  20 &  1--Apr--2001   &  1974  &  6.69  &   3.05  &  2.5    &   Kim et al. (2012) \\
  21 &  15--Jun--2000  &  1964  &  7.11  &   2.49  &  2.2    &   Kim et al. (2012) \\
  22 &  29--Aug--2005  &  2033  &  7.5   &   1.39  &  2.7    &   You et al. (2012) \\
  23 &  14--Jan--2002  &  1985  &  7.78  &   3.90  &  2.3    &   Kim et al. (2012)  \\
  24 &  4--May--2000   &  1962  &  7.92  &   1.78  &  1.8    &   Kim et al. (2012) \\
  25 &  25--Jul--1999  &  1952  &  8.50  &   1.84  &  3.4    &   Kim et al. (2012) \\
  26 &  5--May--2000   &  1962  &  9.06  &   1.36  &  2.1    &   Kim et al. (2012) \\
  27 &  30--Aug--2006  &  2047  &  9.9   &   0.85  &  0.92   &   You et al. (2012) \\
  28 &  14--Jan--2002  &  1985  & 10.06  &   2.33  &  1.7    &   Kim et al. (2012) \\
  29 &  15--Jun--2000  &  1964  & 10.31  &   1.18  &  1.1    &   Kim et al. (2012) \\
  30 &  4--Apr--2000   &  1961  & 11.46  &   1.09  &  2.0    &   Kim et al. (2012) \\
  31 &  25--Jul--1999  &  1952  & 11.7   &   0.97  &  2.1    &   Kim et al. (2012) \\
  32 &  14--Jan--2002  &  1985  & 12.16  &   1.59  &  1.5    &   Kim et al. (2012) \\
  33 &  15--Jun--2000  &  1964  & 12.27  &   0.83  &  0.6    &   Kim et al. (2012) \\
  34 &  4--May--2000   &  1962  & 12.64  &   0.7   &  0.8    &   Kim et al. (2012) \\
  35 &  28--Dec--2001  &  1984  & 14.89  &   0.91  &  1.6    &   Kim et al. (2012) \\
  36 &  4--Apr--2000   &  1961  & 15.33  &   0.609 &  1.0    &   Kim et al. (2012) \\
  37 &  5--Apr--2008   &  2068  & 120.0  &   0.003 &  0.017  &   Poomvises et al. (2012) \\
\hline
 \end{tabular}
 \end{table}

As seen from the Table~\ref{tab1}, the derived magnetic-field values are consistent
with other estimates in a similar distance range.
In Figure~\ref{magexmp} the model calculated magnetic fields at 3 Rs, 5 Rs, 6 Rs, 7 Rs, 10 Rs, 12 Rs
are shown and compared to that measured. Red points denote the directly measured IMF (from Table~\ref{tab1}) that are close to these distances.
So, we conclude that our model adequately describes the observed magnetic fields.
The comparison of the calculated magnetic field with the observations is
complicated by the fact that many of the observations made using different
methods and using different solar corona plasma density models.
Moreover, large-scale and small-scale irregularities, as well as flows of dense and rarefied plasma move with different speeds in the solar corona and interplanetary space, creating regions of high and sparse density at different distances from the Sun.
Even a cursory examination of the measured magnetic fields (Table~\ref{tab1}, column 6) catalogued from
the literature (Table~\ref{tab1}, column 7) shows a wide disparity between the
determinations of various authors. In spite of the large scatter in the
observations, we conclude that they are adequately reproduced by
our model. Such a difference can also result from either calibration differences
or the influence of different solar activity events.
Observed and model magnetic fields differ also because it is hard to distinguish the
enhanced magnetic fields, for example in a CME or a streamer region, from the
background magnetic field.
Our model determines the background magnetic-field strength radial distribution
and cycle evolution.
The difference in the measured magnetic fields can also result from the
dependence of magnetic fields from the selected place of observed event.
\cite{Hariharan2014} estimated the magnetic-field strength in the solar corona
ahead of and behind the MHD shock front associated with a
CME at a distance of $\approx$ 2 Rs. They found magnetic fields of $\approx$ (0.7\,--\,1.4)
$\pm$0.2 G and $\approx$ (1.4\,--\,2.8) $\pm$0.1 G respectively.
\cite{Bemporad2010} determined the plasma parameters of a fast
CME-driven shock associated with the solar eruption of
2002 March 22. According to their study, the magnetic field undergoes
a compression from a pre-shock value of $\approx$0.02 G up
to a post-shock magnetic field of $\approx$0.04 G.

\begin{table}
\caption{Magnetic fields observed by Helios 1 (H1) and Helios 2 (H2), and that calculated using
our model at the same distances from the Sun and the same time.}
\label{tab2}
\tabcolsep=0.6em
\begin{tabular}{llcccccc}
\hline
  N &   space- & Peri-                & CR     & R              &   R        &      B     &     B              \\
     &   craft    &  helios              &           & R$_{AU}$  &  (Rs)     &    calc.  &    obs.           \\
     &             &                         &           &                  &             &    (nT)   &    (nT)            \\
\hline
   1 &      2    &    3                     &    4   &      5       &   6         &       7     &       8             \\
\hline

  1  &  H2  &   2--Nov--1978      &  1674  &   0.354  &   76.11   &  31.17    & 21.9$\pm$10.3   \\
  2  &  H2  &  30--Apr--1978     &  1667  &   0.356  &   76.54    &  28.39    & 27.0$\pm$12.5   \\
  3  &  H2  &  26--Oct--1977     &  1660  &   0.362  &   77.83    &  17.97    & 17.7$\pm$7.8    \\
  4  &  H2  &  23--Apr--1977     &  1654  &   0.372  &   79.98    &  14.86    & 19.7$\pm$11.1   \\
  5  &  H1  &  29--Apr--1978     &  1667  &   0.373  &   80.195   &  25.86   & 24.3$\pm$10.1   \\
  6  &  H2  &   9--Nov--1979      &  1688  &   0.375  &   80.625   &  34.03   & 20.2$\pm$11.2   \\
  7  &  H1  &  13--Apr--1977     &  1653  &   0.381  &   81.915   &  14.5     & 18.3$\pm$9.9    \\
  8  &  H1  &  21--Oct--1977     &  1660  &   0.387  &   83.205   &  15.72   & 15.7$\pm$9.7    \\
  9  &  H2  &   7--May--1979     &  1681  &   0.389  &   83.635   &  30.62   & 24.9$\pm$11.5   \\
  10 &  H1  &   5--Nov--1978     &  1674  &   0.394  &   84.71     &  25.16   & 18.5$\pm$11.1   \\
  11 &  H1  &  21--Nov--1979    &  1688  &   0.394  &   84.71     &  30.83   & 17.0$\pm$8.7    \\
  12 &  H2  &  19--Oct--1976    &  1647  &   0.396  &   85.14     &  11.31   & 18.2$\pm$11,7   \\
  13 &  H1  &  29--May--1980   &  1695  &   0.398  &   85.57     &  25.45   & 17.9$\pm$11.3   \\
  14 &  H1  &   5--Oct--1976     &  1646  &   0.406  &   87.29     &  10.10   & 16.6$\pm$10.9   \\
  15 &  H1  &  14--May--1979   &  1681  &   0.418  &   89.87     &  26.52   & 21.0$\pm$11.6   \\
  16 &  H1  &   5--Jan--1980     &  1702  &   0.438  &   94.17     &  25.54   & 18.9$\pm$10.6   \\
  17 &  H1  &  13--Jun--1981    &  1709  &   0.469  &  100.835   &  19.35   & 16.0$\pm$10.6   \\
\hline
 \end{tabular}
 \end{table}

The Helios mission provides one of the best magnetic-field observations.
The observations performed by Helios 1 and Helios 2 spacecraft allow one
to monitor the IMF conditions in the inner heliosphere from 0.3 to 0.6 AU.
We also compare the calculated magnetic-field strength (Equation~\ref{fmag})
with in-situ measurements made by the Helios 1 and Helios 2 spacecraft (\opencite{Villante1982}).

In Table~\ref{tab2}, the hourly averages of the observed unsigned IMF radial
component measured on Helios 1 (H1) and Helios 2 (H2) (column 2)
spacecraft are presented (column 8, from \cite{Villante1982}).
Magnetic fields in column 7 were calculated using the model (Equation~\ref{fmag})
at the same distances (columns 5, 6)
and time (columns 3, 4) as it was made in \cite{Villante1982}.
Model calculated magnetic fields at 75 Rs, 80 Rs, 85 Rs, 90 Rs, 95 Rs, and 100 Rs
are shown in Figure~\ref{maghelios}.
Red points denote the IMF measured by Helios 1 and Helios 2 spacecraft from Table~\ref{tab2}.
The agreement between our model predictions and that derived from direct,
in-situ measurements is quite satisfactory.
From a comparison of Figures~\ref{mag}, \ref{magexmp}, \ref{maghelios},
\ref{magr} and Tables~\ref{tab1}, \ref{tab2}
of different magnetic-field data as observed and as calculated
from our model  (Equation~\ref{fmag}), we conclude that our model adequately explains
the major features of the magnetic fields at different distances from the Sun
during different cycle phases.
Therefore, the comparison of the observed magnetic fields with those predicted by our model
shows that the magnetic field in the heliosphere is determined by the cycle variations of
the sum of poloidal and toroidal components of the solar global magnetic field.

\section{Coronal Magnetic Field Models}  \label{secmagmod}

\begin{figure}
   \centerline{\includegraphics[width=1.\textwidth,clip=]{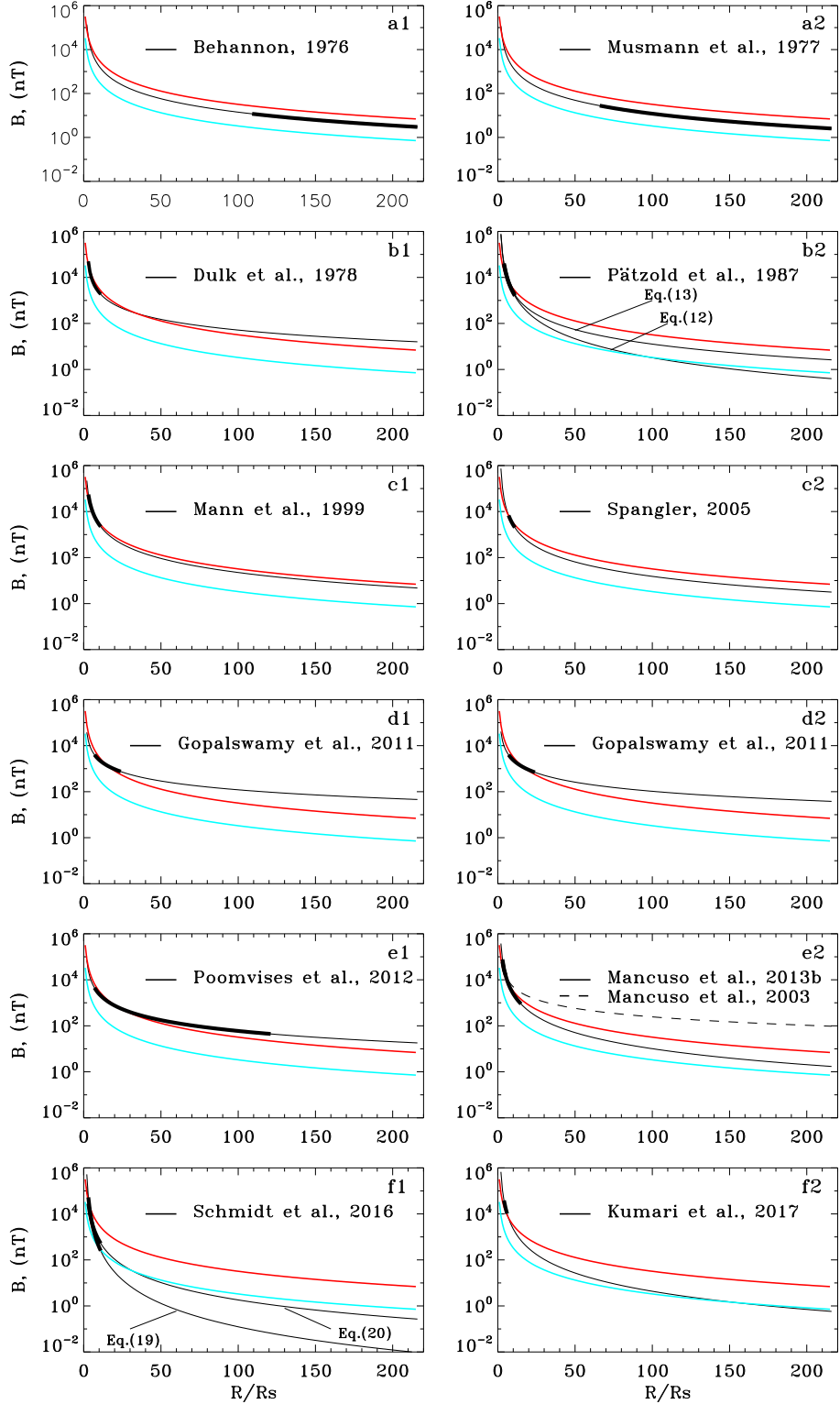}}
      \caption{Magnetic field distributions from 1 Rs to 1 AU calculated for different models (thin black profiles).
               Thick black lines denote the part of the profiles marking the distances for which the
               models were created by their authors.
               Red profiles denote the distribution of magnetic fields
               using our model for the maximum of Cycle 21 (CR 1712).
               Blue profiles denote the distribution of magnetic fields
               using our model for the minimum of Cycle 24 (CR 2079).}
   \label{models}
 \end{figure}

Magnetic field measurements are important for testing
coronal magnetic-field models. Magnetic-field models have played a very
important role in the interpretation of different solar activity phenomena and in
the study of outward flowing coronal material into the interplanetary space.
In this Section, first we summarize various models of deriving the magnetic-field
radial distribution. Models discussed below are presented in the form of functions in Figure~\ref{models}.
Assuming the conservation of the magnetic flux in the interplanetary space
the magnetic field can be continued to arbitrary radial distances $r$ from the Sun (\opencite{Parker1958}) by

\begin{equation}
B(r) = B_s \times \left(\frac{R_s}{r}\right)^2
\label{fb}
\end{equation}

\noindent where $B_s$ is the magnetic field at the photosphere,
obtained using some measurements and assumptions  (e.g. \opencite{Mann1999})
or the magnetic field $B_s$ can be composed of the field of an active
region and that of the quiet Sun (\opencite{Warmuth2005}).
$Rs$ is the radius of the Sun;
$r$ is the distance from the center of the Sun in units of the solar radius.

Magnetic-field radial profile can also be represented by an empirical formula of the form

\begin{equation}
B(r) = K \times r^{-\alpha}
\label{fk}
\end{equation}

\noindent where $K$ is a coefficient derived from some approximation of different measurements or models.
$r$ is the distance from the center of the Sun in units of the solar radius.
A significant number of models were created based on the relation~\ref{fk}.
(e.g. \opencite{Dulk1978}; \opencite{Patzold1987}; \opencite{Mancuso2003};
\opencite{Gopalswamy2011}). But a large number of models have used the
Equation~\ref{fk} formally. The magnetic field values are not included in such formulas.

Traditionally, the problem of a radial magnetic-field distribution has been solved by
combining the data obtained using different observations of magnetic fields measured at different
times by different methods at different distances from the Sun and on different spacecraft.
Then the data are summarized on one curve and fitted by a function like the Equations~\ref{fb}, \ref{fk}.
Such models depend on $r$ only. They do not reflect the actual distribution
and cycle evolution of magnetic fields.

Figure~\ref{models} summarizes radial magnetic-field distributions calculated using
different models. The profiles were calculated for radial distances
from 1 Rs to 1 AU (thin black profiles).
Thick black  lines denote the part of the profiles marking the distances for which the
models were created by their authors.
For comparison, the radial magnetic-field profiles calculated
using our model are also shown in Figure~\ref{models}.
The magnetic-field radial profiles corresponding to the maximum
and minimum solar cycle phases are selected.
Red lines denote the radial distribution of magnetic fields calculated
using our model for the maximum of Cycle 21 (CR 1712).
Blue lines denote the radial distribution of magnetic fields calculated
using our model for the minimum of Cycle 24 (CR 2079).
Such model calculated magnetic-field profiles can be used as a real
magnetic field limitations for different models.

Using measurements of the IMF from several
different spacecraft, \cite{Behannon1976} showed that the radial component
of magnetic field between 0.5 AU and 5 AU can be fitted by a function

\begin{equation}
B(r) = 3.0 \times 10^{-5} \left(\frac{r}{216}\right)^{-2}
\label{fbehannon1976}
\end{equation}

\noindent where $B(r)$ is in G, $r$ is the distance from the center of the
Sun in units of the solar radius (Figure~\ref{models}a1).

\cite{Musmann1977} derived the similar expression for
radial variation of the IMF between 0.3 AU and 1.0 AU
from Helios 1 data during solar minimum

\begin{equation}
B(r) = 1.18 \times 10^{-4} r^{-2}
\label{fmusmann1977}
\end{equation}

\noindent where $B(r)$ is in T, $r$ is in units of the solar radius (Figure~\ref{models}a2).

One of the most often used model is the empirical formula proposed by \cite{Dulk1978}.
\cite{Dulk1978} concentrated their attention on the magnetic fields above active regions.
Using different techniques, and different observational data, they proposed
an empirical single parameter formula for magnetic-field radial profile calculation from 1.02 to 10 r/Rs

\begin{equation}
B(r) = 0.5 \times \left(\frac{r}{Rs} - 1\right)^{-1.5}
\label{fdulk1978}
\end{equation}

\noindent where $B(r)$ is in G, $R_s$ is the solar radius,
$r$ is in units of the solar radius (Figure~\ref{models}b1).
They pointed out that the magnetic field in the corona can vary from one active region to another by
an order of magnitude and that the Equation~\ref{fdulk1978} is consistent with the different data used to
within a factor of about 3.

\cite{Patzold1987} derived the mean coronal magnetic field from Faraday rotation measurements
for $3 \le r \le 10$ Rs during solar minimum in 1975\,--\,1976 (Figure~\ref{models}b2, Eq.(12)).

\begin{equation}
B(r) = 7.9  \times 10^{-4}  r^{-2.7}
\label{fpatzold1987}
\end{equation}

The magnetic-field profile derived using a fit to the Faraday
rotation data with a dipole term and an interplanetary term is given
by an expression (Figure~\ref{models}b2, Eq.(13))

\begin{equation}
B(r) = (6r^{-3} + 1.18r^{-2})\times 10^{-4}
\label{fpatzold1987}
\end{equation}

\noindent where $B(r)$ is in T, $r$ is in units of the solar radius.
Clearly our model profile is closer to the \cite{Patzold1987} Eq. 13
profile than to the \cite{Patzold1987} Eq. 12 profile.
\cite{Patzold1987} Eq. 12 profile agrees well with our model profile for solar
maximum at shorter distances, but it is lower than our model profile for solar
minima.

\cite{Mancuso2000} and \cite{Spangler2005} proposed a method for deriving
the strength and spatial structure of the solar coronal magnetic
field using the observations of the Faraday rotation of radio sources (radio galaxy)
occulted by the solar corona at heliocentric distances of 6\,--\,10 Rs (Figure~\ref{models}c2)

\begin{equation}
B(r) = 0.06\left(\frac{r}{R_1} \right)^{-3} + 3.1\left(\frac{r}{R_1} \right)^{-2}
\label{fspangler2005}
\end{equation}

where $B(r)$ is in nanoTesla, $R_1$ is one astronomical unit,
$r$ is the distance from the Sun in units of the solar radius. In this
model, the field changes polarity at the coronal neutral line \cite{Spangler2005}.

Several models of the magnetic-field radial distribution were based on CME observations.
\cite{Gopalswamy2011} determined the coronal magnetic field strength using
white-light coronagraph measures of the shock standoff distance and the radius of curvature
of the flux rope during the 2008 March 25 CME.
They showed that the radial profile of the magnetic field can be
represented by a power law of the form

\begin{equation}
B(r) = p \times r^{-q}
\label{fgopals2011}
\end{equation}

\noindent where $B(r)$ is in G, with $r$ in units of Rs. $p$ and $q$ are the coefficients that
depend on the plasma density model.
Using \cite{Saito1977} density model with $\gamma=4/3$ and for distances $> 9$ Rs they
got $p = 0.377$ and $q = 1.25$ (Figure~\ref{models}d1).
Using \cite{Leblanc1998} density radial-distribution model they got $p = 0.409$ and $q = 1.3$ (Figure~\ref{models}d2).
\cite{Gopalswamy2011} model extrapolation results in a slightly flatter magnetic-field
profiles compared to that from our model.
\cite{Gopalswamy2011} model magnetic-field profiles agree with our model profiles at shorter distances,
but the difference increases with the distance.
They determined the coronal magnetic field strength in the heliocentric distance range 6\,--\,23 solar radii.

Following the standoff-distance method,
\cite{Poomvises2012} derived radial magnetic-field strength in the heliocentric distance
range from 6 to 120 Rs using data from Coronagraph 2 and Heliospheric Imager I
instruments on board the Solar Terrestrial Relations Observatory spacecraft.
They found that the radial magnetic-field strength decreases
from 28 mG at 6 Rs to 0.17 mG at 120 Rs.
They derived magnetic-field profiles in the form of Equation~\ref{fgopals2011}.
Using \cite{Saito1977} density mode and  $\gamma=4/3$ they
got $p = 845.870$ and $q = 1.59$. Using \cite{Leblanc1998}
density radial distribution model they
got $p = 706.383$ and $q = 1.54$. The coefficients are close in value,
so the resulting profiles coincide (Figure~\ref{models}e1).

Several models of the magnetic field radial distribution are based on solar radio
emission analysis.
\cite{Mann1999} have shown that coronal EIT waves and coronal
shock waves associated with type II radio bursts can be used to determine
magnetic field in the solar corona. They proposed a model of radial magnetic-field
distribution for quiet regions at solar minimum

\begin{equation}
B(r) = 2.2 \times \left(\frac{Rs}{r}\right)^2
\label{fman1999}
\end{equation}

\noindent where $B(r)$ is in G, $r$ is in units of Rs (Figure~\ref{models}c1).
They determined magnetic field strength between 1.1\,--\,2.1 solar radii.
The model profile matches well our estimates for coronal
magnetic field at solar maxima.

\cite{Mancuso2003} studied coronal plasma analyzing type II radio bursts and SOHO Ultra Violet Coronagraph Spectrometer (UVCS) observations. The data sample comprises 37 metric type II radio bursts observed by ground based radio spectrographs in 1999, during the rising phase of Cycle 23.
The shock speeds were used to set upper limits
to the magnetic field above active regions.
An average functional form of the magnetic-field estimates can be represented
by the following radial profile, valid between about 1.5 and 2.3 Rs

\begin{equation}
B(r) =  (0.6\pm 0.3) \times (r-1)^ {-1.2}
\label{fmancuso2003}
\end{equation}

\noindent where $B(r)$ is in G, $r$ is in units of Rs (Figure~\ref{models}e2).

Coronal magnetic field can be inferred from band-splitting of type II radio bursts (\opencite{Vrsnak2002}).
\cite{Mancuso2013b} have also analyzed the band-splitting of
type II radio burst to determine the coronal magnetic-field strength
in the heliocentric distance range $\approx$1.8\,--\,2.9 Rs.
The same radial profile was obtained higher up in the corona
by \cite{Mancuso2013a} based on Faraday rotation measurements
of extragalactic radio sources occulted by the solar corona.

\begin{equation}
B(r) = 3.76 \times r^{-2.29}
\label{fvrsnak2002}
\end{equation}

\noindent where B(r) is in G, $r$ is in units of Rs (Figure~\ref{models}e2).
The profiles derived describe the magnetic field in a range of
heliocentric distances from 1.8 Rs to 14 Rs.
Clearly our model profile is more in line with the \cite{Mancuso2013b}
profile than with the \cite{Mancuso2003} profile.

Comparison of models developed by \cite{Behannon1976},  \cite{Musmann1977},
\cite{Patzold1987} (Ed.(13)), \cite{Spangler2005}, \cite{Mancuso2013b}
(Figure~\ref{models}a1, a2, b2 (Eq.13), c2, e2)
shows that these model radial profiles and those predicted by our model
for the maxima and minima corona states agrees very well. The profiles lie
between that calculated for the maxima and minima corona states using our model.
The agreement is quite satisfactory, although it is impossible from the profiles to prefer
one model above the other.

\cite{Schmidt2016} derived the profile of the strength of the magnetic field in front
of a CME-driving shock based on white-light images and the standoff-distance method.
They also simulate the CME and its driven shock with a 3-D MHD code.
They found good agreement between the two profiles (within $\pm$30\%) between 1.8 and 10 Rs.
The authors noticed that in their model a magnetic-field profile
is decreasing stronger than a monopolar (wind-like) magnetic-field
profile $\approx r^{-2}$ and a dipolar profile  $\approx r^{-3}$ for an
ideal spherically symmetric system.
Magnetic-field strength profile derived can be represented as

\begin{equation}
log_{10}B_{calc}(r) = (-3.32 \pm 0.5) \times log_{10}(r) + (3.72 \pm 0.5)
\label{fschmidt2016}
\end{equation}

and that simulated with the 3-D MHD code

\begin{equation}
log_{10}B_{sim}(r) = (-2.47 \pm 0.5) \times log_{10}(r) + (3.19 \pm 0.5)
\label{fschmidt2016c}
\end{equation}

\noindent where $B(r)$ is in mG,  $r$ is in units of Rs (Figure~\ref{models}f1)
The profiles are similar to our model minima profile at shorter distances, but the difference
is growing rapidly with increasing distance.

Using radio and white-light observations, \cite{Kumari2017} have shown
that a single power-law fit is sufficient to describe
magnetic fields in the heliocentric distance range $\approx$2.5\,--\,4.5 Rs

\begin{equation}
B(r) = 6.7 \times  r^{-2.6}
\label{fkumari2017}
\end{equation}

\noindent where B(r) is in G, $r$ is in units of Rs (Figure~\ref{models}f2).
Resulting profile matches our model profile for magnetic-field maximum
at shorter distances and it close to our model profile for magnetic-field minimum at large distances.

From Figure~\ref{models} it is seen, that magnetic-field profiles derived using
different models are different.
The difference for different models in the variation in
magnetic field at 1 AU is an order of magnitude.
The main difference between our model and the models described above
is that they give only one value for one point at a certain distance from the Sun.
They don't take into consideration the solar cycle magnetic-field evolution.
Our model gives more realistic magnetic-field radial distributions taking
into account the cycle variations of the solar toroidal and poloidal magnetic fields.

\section{Discussion} \label{secdiscussion}

In this study a new model has been presented for the radial $B_x$ component of the IMF
calculation at various distances from the Sun during different solar cycle phases.
Results show a rather good match between the measured $B_x$ component of the IMF and the model
predictions. The magnetic fields from WSO were used in the study. But it is known that
the magnetic-field measurements at different observatories are different.
To compare data from different observatories, synoptic maps from Wilcox Solar Observatory (WSO), Mount Wilson Observatory (MWO), Kitt Peak (KP), SOLIS, SOHO/MDI, and SDO/HMI measurements of the photospheric field were analyzed by \cite{Riley2014}, \cite{Virtanen2016}, \cite{Virtanen2017}.
The comparison has shown that while there is a general qualitative agreement
in the measured data, there are also some significant differences (\opencite{Riley2014}).
The observatories give a similar overall view of the solar magnetic-field and the heliospheric
current sheet evolution over four last cycles. However, there are some periods
when the data disagree with each other (\opencite{Virtanen2016}).
The differences between the data sets can be due to
instrument problems, the choice of the spectral lines that may be formed
at different heights and may not measure the same magnetic field,
the measurement and treatment of polar fields,
for ground-based instruments, atmospheric turbulence can
significantly degrade the image quality,
the algorithms used to create synoptic maps (\opencite{Riley2014}).
So, the assumption that high-resolution maps from one observatory
can be transformed to a lower-resolution map of another by simple averaging is strictly not true.
Scaling factors are needed when comparing
synoptic maps from different observatories.
Therefore, the conversion factors were computed by \citeauthor{Riley2014}, \citeyear{Riley2014}
that relate measurements of different observatories using both synoptic map pixel-by-pixel and
histogram-equating techniques.
\cite{Virtanen2017} also proposed a method for scaling the photospheric magnetic
fields based on the harmonic expansion. The benefit of the harmonic scaling method is that it can be
used for data sets of different resolutions.

WSO provides the longest and most homogenous magnetic-field observations,
a lot of investigations based on WSO data.
But WSO provides the lowest values for photospheric magnetic field data, and thus a
weaker coronal magnetic-field strength, than the Mount Wilson Observatory, Kitt Peak, SOLIS,
SOHO/MDI, and SDO/HMI (\opencite{Riley2014}).
The size of a WSO pixel is 3 minutes of arc in sky coordinates, or 180 arc sec,
which at disk-center represents about 126 Mm, or about 10 degrees of heliographic longitude.
Such large pixels cannot resolve active regions so most active region flux is not detected by WSO.
So, the WSO's very low spatial resolution can be a part of the explanation of the good match,
because it was found that the origin of IMF is the large-scale photospheric magnetic fields
(\opencite{Ness1964}; \opencite{Ness1965}; \opencite{Ness1966}; \opencite{Severny1970}).
IMF evolves in response to the solar coronal and photospheric magnetic fields at its base.
The source of the interplanetary-sector structure is associated with large-scale photospheric magnetic-field patterns, the patterns of weak photospheric background fields (\opencite{Wilcox1968}, \opencite{Wilcox1967}) with the area equal to about one-fourth the area of the solar disk (\opencite{Scherrer1972}),
or according to Plyusnina \citeyear{Plyusnina1985} the size of unipolar photospheric magnetic-field pattern is on the average larger than $40^{\circ}$.
Therefore, such a good coincidence between the measured magnetic field
and that predicted by the model is largely because of WSO's low-resolution
measurements. The active region total flux is unresolved. The WSO measured magnetic
field is the large-scale field that form and govern IMF.

It should be noted, that the photospheric magnetic fields and IMF
are measured from the Earth and Earth's orbit.
Only flux from different latitudes that is detected at the Earth is considered.
This does not imply that approximately
the same proportion of low- and high-latitude fields contribute to the IMF $B_x$ component.
We don't see the real radial polar magnetic field from the Earth.
Since the value of the polar magnetic field observed from the Earth is used, not the full magnetic flux of the polar regions is really measured, but only the tangential component
of the polar magnetic field of the visible polar area.
The contribution of polar regions is far less than that of low-latitude fields especially during
maxima phases, when the polar region influence is negligible despite the fact that
the polar flux is much more unipolar than the low-latitude flux over most of the cycle.
The role of polar fields increases during minima phases.
The central part of the visible solar disc contributes the major amount of magnetic
flux to the magnetic field measured at the Earth.
Scherrer et al. (\citeyear{Scherrer1977}) have shown that about half the contribution
to the mean (sun-as-a-star) field comes from the center 35\% of the disc area.

The active region magnetic-field influence on the IMF is insignificant beyond source surface. It was found that the origin of IMF is the large-scale photospheric magnetic fields. The source of the interplanetary-sector structure is associated with the pattern of weak photospheric background fields (\opencite{Wilcox1968}). The background solar magnetic field is represented predominantly by a radial field (as seen from vectormagnetograms and EUV images). It should be stressed, that it does not show significant variations with latitude. The independence of latitude was also observed by Ulysses in the radial component of IMF (\opencite{Forsyth1996}; \opencite{Smith1995};  \opencite{Balogh1995}). But, as already mentioned earlier, both the photospheric magnetic field and IMF are measured from the Earth and the Earth's orbit. Hence, the contribution of the unipolar regions to the total magnetic flux that reaches the Earth is also mainly from the central solar region.

Finally, it should be noted that the photospheric magnetic fields
must be accurate determined as they are used to calculate coronal magnetic fields and IMF.
So, the magnetic-field measurements from other observatories
can be used with appropriate scaling factor to calculate the IMF at various distances
from the Sun during different cycle phases.

\section{Conclusion} \label{secconclusion}

A new model has been proposed for magnetic field determination at different
distances from the Sun throughout solar cycles.
The model depend on the observed large-scale photospheric magnetic fields.
The direct observations of
the large-scale non-polar photospheric ($\pm55^{\circ}$) and
polar  (from $55^{\circ}$ N to $90^{\circ}$ N and
from $55^{\circ}$ S to $90^{\circ}$ S) magnetic fields were used,
which are the visible manifestations of cyclic changes in the
toroidal and poloidal components of the solar global magnetic field.
The model magnetic field is determined as the sum of the non-polar photospheric
magnetic field (toroidal component of the solar global magnetic field)
and the polar magnetic field (poloidal component) cycle variations.

The agreement between our model predictions and magnetic fields derived from direct,
in-situ, measurements at different distances from the
Sun and at different time is quite satisfactory.
From a comparison of the magnetic fields as observed  and as calculated from our
model at 1 AU, we also conclude that the model magnetic-field strength variation adequately explains
the major features of the IMF $B_x$ component cycle evolution at the Earth's orbit.
The model CR-averaged calculated magnetic fields correlate with
CR-averaged IMF $B_x$ component at the  Earth's orbit with a coefficient of 0.688.
Correlation between seven CR-averaged calculated magnetic fields
with IMF $B_x$ component reaches 0.808.
For Cycles 21\,--\,24, the model CR-averaged calculated magnetic fields
correlate with CR-averaged IMF $B_x$ component at the Earth's orbit
with a coefficients of 0.42, 0,69, 0.73, and 0.74 and that for seven CR-averaged values
with a coefficients of 0.52, 0,88, 0.85, and 0.88

The magnetic-field cycle variations derived for different distances
should be regarded as a good approximation of the radial cycle behavior of the
magnetic fields in the heliosphere. Thus, this method of the magnetic-field
calculation is a useful tool for the description of solar corona and interplanetary
large-scale magnetic-field cycle evolution at differen distances from the Sun.
So, the model should be regarded as a good approximation
of the cycle behavior of the magnetic field in the heliosphere.

Magnetic-field profiles derived from our model are similar
to those of empirical models and previous estimates.
The major difference between our model and the models described above
is that they give only one value for one point at a certain distance from the Sun.
They don't take into consideration the solar cycle magnetic-field evolution.
Our model gives more realistic magnetic-field radial distributions taking
into account the cycle variations of the solar toroidal and poloidal magnetic fields.

A particularly interesting finding has been the decrease in
maximum of the photospheric magnetic-field magnitudes from Cycle 21 to Cycle 24.
It should be noted that the minimum values of the magnetic field in the
solar activity minima are also decreased from Cycle 21 to Cycle 24.
The polar magnetic field also decreased.
Such changes in magnetic fields reflect the
time evolution of the toroidal and poloidal components of the solar global magnetic field. It means that
both components and therefore, the solar global magnetic field decreased from Cycle 21 to Cycle 24.

%
 \begin{acks}

The author express the appreciation to the anonymous referee
for a very thorough and helpful review of the paper.

Wilcox Solar Observatory data used in this study was obtained via
the web site
$http://wso.stanford.edu \, at \, 2018:03:11\, 01:13:34$ PST
courtesy of J.T. Hoeksema. The Wilcox Solar Observatory
is currently supported by NASA.

Data on IMF were obtained from multi-source OMNI 2 data base via the web site
$https://omniweb.gsfc.nasa.gov/ow.html$. The author thanks the GSFC/SPDF and OMNIWeb
for the opportunity to use this data.

\end{acks}

%
%
\bibliographystyle{spr-mp-sola}

 \bibliography{bilenko}
%
%
%
%

\end{article}
\end{document}